# Size-dependent bandgap and particle size distribution of colloidal semiconductor nanocrystals

Diego L. Ferreira[1]*, J. C. L. Sousa[2], R. N. Maronesi[1], J. Bettini[3], M. A. Schiavon[2], Alvaro V. N. C. Teixeira[1], and Andreza G. Silva[1]

[1]*Departamento de Física, CCE, Universidade Federal de Viçosa, Avenida P. H. Rolfs, Viçosa, MG, 36570-900, Brazil.*

[2]*D Departamento de Ciências Naturais, Universidade Federal de São João Del Rei, Campus Dom Bosco, 36301-160, São João del-Rei, MG, Brazil..*

[3] *Laboratório Nacional de Nanotecnologia, Centro Nacional de Pesquisa em Energia e Materiais, 13083-970, Campinas, SP, Brazil.*

A new analytical expression for the size-dependent bandgap of colloidal semiconductor nanocrystals is proposed within the framework of the finite-depth square-well effective mass approximation in order to provide a quantitative description of the quantum confinement effect. This allows one to convert optical spectroscopic data (photoluminescence spectrum and absorbance edge) into accurate estimates for the particle size distributions of colloidal systems even if the traditional effective mass model is expected to fail, which occurs typically for very small particles belonging to the so-called strong confinement limit. By applying the reported theoretical methodologies to CdTe nanocrystals synthesized through wet chemical routes, size distributions are inferred and compared directly to those obtained from atomic force microscopy and transmission electron microscopy. This analysis can be used as a complementary tool for the characterization of nanocrystal samples of many other systems such as the II-VI and III-V semiconductor materials.

## I. INTRODUCTION

Motivated by Ekimov's first experimental observation of the size dependence of nanocrystal optical properties in semiconductor-doped glasses[1,2], Efros and Efros conducted pioneering theoretical investigations of quantum confinement effects in semiconductor spherical microcrystallites[3]. In the framework of the effective mass approximation for the confined charge carriers, interband optical absorption coefficients were calculated in two limiting cases or the so-called quantum confinement regimes, depending on the ratio of the crystallite radius ($R$) to the effective Bohr radius of the electron-hole pair ($a_B$): the strong confinement limit ($R/a_B \ll 1$, individual particle confinement regime) and the weak confinement limit ($R/a_B \gg 1$, exciton confinement regime). An intermediate confinement regime was also introduced for $a_h \ll R \ll a_e$ ($a_h$ and $a_e$ are the Bohr radii of the hole and the electron, respectively). Expressions for the energy of the first excited electronic state were derived for each case so that the bandgap enlargement due to size quantization effects (the bandgap of the semiconductor particle relative to the bulk value) could be first estimated.

Since Efros and Efros seminal contribution[3], several models have been proposed to understand the size-dependent bandgap of low dimensional semiconductor structures especially in the size range of small particles corresponding to the strong confinement regime ($R/a_B \ll 1$). However, development of a theoretical analytical model suitable for quantitative predictions is still a partially solved problem.

One of the most used theoretical models that allow a relatively simple analytical relationship between bandgap and particle size is the much quoted Brus model[4-6]. In its simplest form, the widely known Brus equation results from an effective mass model for spherical particles in the case of strong size quantization. As an improvement to Efros and Efros treatment[3] of the strong confinement regime, the coulomb interaction between electron and hole was included by means of first order perturbation theory. Quantum confinement effects on ionization potentials, electron affinities and redox potentials were then analyzed in details in the sense of Brus model. The blue shift of the absorption spectrum was also obtained in reasonable agreement with experiment for large clusters[7]. However, experimental observations carried out extensively have revealed that in a system composed by extremely small nanocrystals ($R$ as small as 1-2 nm), near the so-called strong confinement limit, the observed bandgap shift with respect to the bulk value is much smaller than the theoretical prediction[8-10]. Consequently, in the size range corresponding to the strong confinement regime, Brus equation fails to fit the empirical sizing curves (nanocrystal bandgap vs size) published by several groups by combining experimental data for different materials[11-13]. In the specific case of the size distributions analyzed in reference[13] for various samples of ZnO nanocrystals, the particle size obtained from absorption onset measurement and Brus sizing curve deviates roughly by 25% from the maximum of the corresponding transmission electron microscopy histogram. Such discrepancy has been attributed mainly to the boundary constraint of the infinite barrier model, which constitutes the underlying assumption for the main results of Efros and Efros[3], and Brus[5]. In this context, Kayanuma and Momiji[14] introduced variational calculations of the ground state energy of an electron-hole pair system confined in a microsphere by finite potential barriers. It was shown that the effect of relaxation of the boundary constraint is quite significant and must be taken into account to analyze the experimental data properly. Other researchers[15-17] adopted a more refined method based on the finite-depth square-well effective mass approximation and suitable for quantitative predictions. Assuming a spherical finite potential well, electron and hole energies can be estimated numerically by solving appropriate nonlinear algebraic eigenvalue equations. Nanda *et al.*[15] and Pellegrini *et al.*[16] investigated systematically the application of this approach to several semiconductor nanocrystals embedded in different matrices and the model

\* To whom correspondence should be addressed:
E-mail: diegolourenconi@gmail.com.

predictions for wide-bandgap semiconductors turned out to be quantitatively accurate.

In addition to the reported theoretical investigations, empirical calibration curves have also been proposed for CdS, CdSe and CdTe colloidal nanocrystals providing useful relationships between the mean size of the nanocrystals and the position of the first excitonic absorption peak[11]. Such empirical functions agree very well with the calculated absorption spectra using time-dependent density functional methods for similar cadmium chalcogenides[18]. A good agreement is also found when an atomistic semiempirical pseudopotential approach is used for calculating the size dependent exciton transition energies of small CdSe nanocrystals[19].

In this paper, a new analytical relationship between the bandgap of a spherical semiconductor nanocrystal and its characteristic size is presented as an alternative to the referred numerical approaches and also to Brus equation in a specific size range ($R/a_B \ll 1$) where this asymptotic formula fails to describe experimental observations (the strong confinement limit). Relevant corrections to the lowest excited state of these quantum confined systems were compiled in order to provide realistic sizing curves (nanocrystal bandgap vs radius). From a simple spectroscopic analysis based on optical absorption and photoluminescence measurements and applied to CdTe colloidal nanocrystals, particle size distributions were estimated and compared directly to those obtained from atomic force microscopy (AFM) and transmission electron microscopy (TEM).

## II. THEORY

### A. Size-dependent bandgap of colloidal semiconductor nanocrystals

Leyronas and Combescot[20] derived analytical expressions for the single particle confinement energies in a spherical nanocrystal with finite potential barriers in order to reproduce impressively well the numerical solutions of the characteristic transcendental eigenvalue equation for any level, barrier height, and confinement size. From them, we can propose in the present paper the exact ground-state wave function for the charge carriers in a spherically symmetric finite potential well with radius $R$:

$$\phi_{v_i}(x_i) = \frac{1}{\sqrt{2\pi R}} \frac{1}{\pi f(v_i) j_1[\pi f(v_i)]} \frac{\sin\left[\frac{\pi}{R} f(v_i) x_i\right]}{x_i}, \quad (1)$$

where $x_i$ is the radial coordinate for the electron ($i = e$) and the hole ($i = h$), $j_1[\pi f(v_i)]$ is a first-order spherical Bessel function with argument $\pi f(v_i)$, and

$f(v_i) = \left[1 + \frac{1}{v_i} + \frac{\left(\frac{\pi}{2}-1\right)^2}{v_i(v_i-1)}\right]^{-1}$ is a quantity defined in terms of the dimensionless parameter $v_i = \left(\frac{V}{\hbar^2/2m_i R^2}\right)^{1/2}$. This finite confining parameter relates the barrier height $V$ and the confinement energy of the charge carrier $i$, characterized by the effective mass $m_i$. The infinite potential limit is reached when $v_i \to \infty$. Assuming that the individual motions of the electron and the hole are strongly quantized in all spatial directions, in accordance with the regime of sufficiently small nanocrystals ($R/a_B \ll 1$), the exciton ground-state wave function $\psi_{v_e,v_h}(x_e, x_h)$ can be factorized into a simple product of the 1S single-particle wave functions $\phi_{v_i}(x_i)$, so that $\psi_{v_e,v_h}(x_e, x_h) \cong \phi_{v_e}(x_e) \times \phi_{v_h}(x_h)$. The energy corresponding to the first excitonic transition or, equivalently, the bandgap of a semiconductor nanocrystal [$E_g(R)$] relative to the bulk value ($E_g^{bulk}$) becomes:

$$E_g(R) = E_g^{bulk} + \frac{\hbar^2}{2m_e R^2}\left[\frac{\pi}{1 + \frac{1}{v_e} + \frac{\left(\frac{\pi}{2}-1\right)^2}{v_e(v_e-1)}}\right]^2$$

$$+ \frac{\hbar^2}{2m_h R^2}\left[\frac{\pi}{1 + \frac{1}{v_h} + \frac{\left(\frac{\pi}{2}-1\right)^2}{v_h(v_h-1)}}\right]^2 \quad (2)$$

$$+ \Delta E_{e-h}(R, v_e, v_h, \varepsilon_s) + \Delta E_{pol}(R, v_e, v_h, \varepsilon),$$

where the second and the third terms correspond to the confinement energies of the electron and of the hole, respectively, in a finite spherical potential well. The fourth term is due to the screened Coulomb interaction between the electron and the hole. It depends explicitly on the nanocrystal radius ($R$), the finite confining parameters for the charge carriers ($v_e, v_h$), and the dielectric constant of the bulk semiconductor material ($\varepsilon_s$). Treating the Coulomb interaction as a first order perturbation to the dominant kinetic energy contribution for small radii, and making use of the Legendre polynomial addition theorem for the $\frac{1}{|\overrightarrow{x_e}-\overrightarrow{x_h}|}$ term, we obtain:

$$\Delta E_{e-h}(R, v_e, v_h, \varepsilon_s)$$
$$\cong \left\langle \psi_{v_e,v_h}(x_e, x_h)\left|-\frac{e^2}{\varepsilon_s|\overrightarrow{x_e}-\overrightarrow{x_h}|}\right|\psi_{v_e,v_h}(x_e, x_h)\right\rangle \quad (3)$$
$$= -\frac{e^2}{\varepsilon_s}(I_1 + I_2),$$

where

$$I_1 = \int d^3 x_h |\phi_{v_h}(x_h)|^2 \int dx_e x_e^2 |\phi_{v_e}(x_e)|^2$$
$$\times \sum_{n=0}^{\infty} \frac{1}{x_h}\left(\frac{x_e}{x_h}\right)^n \Theta(x_h - x_e) \int d\Omega_e P_n(\cos\gamma), \quad (4)$$

and

$$I_2 = \int d^3 x_e |\phi_{v_e}(x_e)|^2 \int dx_h x_h^2 |\phi_{v_h}(x_h)|^2$$
$$\times \sum_{n=0}^{\infty} \frac{1}{x_e}\left(\frac{x_h}{x_e}\right)^n \Theta(x_e - x_h) \int d\Omega_h P_n(\cos\gamma), \quad (5)$$

In Eqs. (4) and (5), $\Theta$ is the usual Heaviside unit step function, $P_n$ is the nth-order Legendre polynomial and $\gamma$ is the angle between the position vectors $\overrightarrow{x_e}$ and $\overrightarrow{x_h}$. The integral of $P_n(\cos\gamma)$ with respect to the solid angle element $d\Omega_i$ for the electron ($i = e$) and the hole ($i = h$) vanishes for all $n \neq 0$: $\int d\Omega_i P_n(\cos\gamma) = 4\pi \delta_{n,0}$. The subsequent integration over the Heaviside function $\Theta(x_i - x_j)$ is performed making use of the identity $\int_0^\infty \Theta(x_i - x_j) g(x_j) dx_j = \int_0^{x_i} g(x_j) dx_j$, where the subscripts $i$ and $j$ are used here to represent different charge carriers and their corresponding radial coordinates, and $g(x_j)$ is a general function of the coordinate $x_j$. All these considerations lead to the following expression for the Coulomb interaction energy [Eq. (3)]:

$$\Delta E_{e-h}(R, v_e, v_h, \varepsilon_s) \cong -\frac{e^2}{\varepsilon_s R}\left(\frac{2\pi^{-\frac{5}{2}}f(v_h)^{-1}f(v_e)^{-\frac{3}{2}}}{j_1[\pi f(v_e)]j_1[\pi f(v_h)]}\right)^2 \left\{-\frac{1}{4}\text{Si}[2\pi f(v_e)] - \frac{1}{8}\text{Si}[2\pi(f(v_h) - f(v_e))] \right.$$
$$\left. + \frac{1}{8}\text{Si}[2\pi(f(v_h) + f(v_e))] + \frac{1}{2}\frac{f(v_e)}{f(v_h)}[\pi f(v_h) - \cos(\pi f(v_h))\sin(\pi f(v_h))]\right\} \quad (6)$$

The expression between braces is written in terms of the $\text{Si}(x)$ sine integral.

The last term in Eq. (2), $\Delta E_{pol}(R, v_e, v_h, \varepsilon)$, is the surface polarization energy which arises from the difference in dielectric constants between the nanocrystal semiconductor material ($\varepsilon_s$) and the surrounding medium ($\varepsilon_m$). As a consequence of this dielectric mismatch, the effective coulomb interaction between the electron and the hole in a spherical semiconductor nanocrystal embedded in a dielectric medium exhibits an additional term caused by the induced surface charge of the sphere[21,22]. From classical electrostatics, Brus derived a polarization potential for a dielectric sphere in the field of a single point charge within it[5]. For one electron-hole pair system, such a potential $[V_{pol}(\vec{x_e}, \vec{x_h})]$ was expressed as a sum of the self-energy of an electron and a hole due its own image charge $[V_s(\vec{x_i})]$ and a mutual polarization contribution coming from the interaction of a carrier with the charge induced by the other one $[V_M(\vec{x_e}, \vec{x_h})]$. Indeed,

$$V_{pol}(\vec{x_e}, \vec{x_h}) = V_s(\vec{x_e}) + V_s(\vec{x_h}) + V_M(\vec{x_e}, \vec{x_h})$$
$$= \sum_{n=0}^{\infty} \frac{e^2 \alpha_n}{2R}\left(\frac{x_e}{R}\right)^{2n} + \sum_{n=0}^{\infty} \frac{e^2 \alpha_n}{2R}\left(\frac{x_h}{R}\right)^{2n} \quad (7)$$
$$- \sum_{n=0}^{\infty} \frac{e^2 \alpha_n}{R}\left(\frac{x_e x_h}{R^2}\right)^n P_n(\cos\gamma)$$

where $\alpha_n$ is defined by $\alpha_n \equiv \frac{(\varepsilon-1)(n+1)}{\varepsilon_s(n\varepsilon+n+1)}$, and $\varepsilon = \varepsilon_s/\varepsilon_m$ is the relative dielectric constant. By assuming infinitely high confining potentials, the dielectric mismatch corrections on excitonic energies in spherical nanocrystals almost cancel each other out and are greatly reduced (in this situation, the contributions from $V_s(\vec{x_e}) + V_s(\vec{x_h})$ and $V_M(\vec{x_e}, \vec{x_h})$ to the potential energy of the electron-hole system have close absolute values and opposite signs). To the best of our knowledge, the combined effect of finite potential barriers and dielectric mismatch on electronic and optical properties of semiconductor nanocrystals has been investigated only in a few works[23-25]. In a very recent publication[25], the dielectric correction for cubic geometry and the eigenstates of the corresponding finite square well were computed for CdTe nanocrystals considering different values of dielectric mismatches and barrier heights. In the present work, in order to account for both dielectric corrections and finite confining potentials in spherically symmetric nanosystems, the electron and hole self-energies and the mutual polarization term from Brus polarization potential [Eq. (7)] were averaged with the proposed exciton ground-state wave function, $\psi_{v_e,v_h}(x_e, x_h)$, for a spherical semiconductor nanocrystal with finite potential barriers, yielding the following analytical expression for the energy shift $\Delta E_{pol}$:

$$\Delta E_{pol}(R, v_e, v_h, \varepsilon) \cong \langle \psi_{v_e,v_h}(x_e, x_h)|V_{pol}(\vec{x_e}, \vec{x_h})|\psi_{v_e,v_h}(x_e, x_h)\rangle = -\frac{e^2}{\varepsilon_s R}\left\{\frac{1}{\pi^2 f(v_e)f(v_h)j_1[\pi f(v_e)]j_1[\pi f(v_h)]}\right\}^2 \quad (8)$$
$$\times \left\{\left(1 - \frac{\sin[2\pi f(v_e)]}{2\pi f(v_e)}\right)g(\varepsilon, v_h) + \left(1 - \frac{\sin[2\pi f(v_h)]}{2\pi f(v_h)}\right)g(\varepsilon, v_e)\right\}$$

where $g(\varepsilon, v_i) = \sum_{n=1}^{\infty} \frac{(\varepsilon-1)(n+1)}{(n\varepsilon+n+1)} \int_0^1 dx\, x^{2n} \sin^2[\pi f(v_i)x]$. A sufficiently high number of terms must be considered in this expansion in order to ensure convergence ($n = 14000$ in our calculations).

Once $\Delta E_{e-h}(R, v_e, v_h, \varepsilon_s)$ and $\Delta E_{pol}(R, v_e, v_h, \varepsilon)$ have been determined from Eqs. (6) and (8), respectively, the bandgap $E_g(R)$ of a semiconductor nanocrystal with respect to the bulk value $E_g^{bulk}$ can be calculated from Eq. (2). For a given system, according to Pellegrini[16] and Nanda[15], the barrier height $V$ entering in the definition of the confining parameters $v_e$ and $v_h$ can be approximated by the difference between the bandgaps of the nanocrystal semiconductor material $E_g^{bulk}$ and of the surrounding medium $E_g^{medium}$, so that $V = [E_g^{medium} - E_g^{bulk}]/2$. The confining potentials for the electron and the hole are assumed to be identical.

At this point, it is worth noting that in the limit of infinite confining potentials ($v_{e,h} \to \infty$ and $f(v_{e,h}) \to 1$), Eqs. (6) and (8) return $\Delta E_{e-h} \to -\frac{4}{\pi}\left\{-\frac{1}{4}\text{Si}(2\pi) + \frac{1}{8}\text{Si}(4\pi) + \frac{\pi}{2}\right\}\frac{e^2}{\varepsilon_s R} = -1.786\frac{e^2}{\varepsilon_s R}$ and $\Delta E_{pol} \to -\frac{2e^2}{\varepsilon_s R}g(\varepsilon, v_i \to \infty)$, respectively. In this limit, the confinement energies (second and third terms in Eq. (2)) exhibit an inverse quadratic dependence on the nanocrystal radius. Therefore, the main result of the well-known Brus model[5] is recovered from the asymptotic form of Eq. (2):

$$E_g \cong E_g^{bulk} + \frac{\hbar^2 \pi^2}{2\mu R^2} - 1.786\frac{e^2}{\varepsilon_s R} + \beta \frac{e^2}{\varepsilon_s R} \quad (9)$$

where
$$\beta = -2g(\varepsilon, v_i \to \infty) = -2\sum_{n=1}^{\infty} \frac{(\varepsilon-1)(n+1)}{(n\varepsilon+n+1)} \int_0^1 dx\, x^{2n} \sin^2(\pi x)$$
and $\mu$ is the reduced electron-hole mass.

## B. Determination of the particle size distribution of colloidal semiconductor nanocrystals

In real systems, regardless the adopted synthesis methods, one has to take into account that there is always a certain distribution of nanocrystal sizes $P(R)$ around a certain mean value. In this context, well established colloidal chemistry approaches combined with post-preparative size-selective precipitation techniques have been able to furnish high quality nanocrystals with size dispersions as narrow as 5%[26,27]. Since the bandgap of a single semiconductor nanocrystal depends strongly on its radius [see Eqs. (2) and (9)], a certain size distribution leads necessarily to a distribution of bandgaps and introduces a pronounced inhomogeneous broadening of the originally discrete resonances in the observed optical spectra. Considering specifically the effect of size nonuniformity on the photoluminescence spectra of semiconductor nanocrystals, the ensemble emission intensity (on the $\lambda$-wavelength scale) can be simulated as[13,28-30]

$$\overline{I_{PL}}(\lambda) = \int_0^\infty N_c(R)\alpha_{ABS}^{(R)}(\lambda_{exc})I_{PL}^{(R)}(\lambda)P(R)dR \quad (10)$$

where $N_c(R)$ is a size-dependent number of carriers available to take part in optical transitions; $\alpha_{ABS}^{(R)}(\lambda_{exc})$ and $I_{PL}^{(R)}(\lambda)$ are the linear absorption coefficient at the excitation wavelength $\lambda_{exc}$ and the emission intensity for a single nanocrystal of radius $R$, respectively; $P(R)$ is the probability distribution function of radii. Assuming that $P(R)$ can be represented either by a normal or by a log-normal dispersion, Eq. (10) has furnished a good fit to experimental photoluminescence data especially for silicon nanoclusters over the size range 2-8 nm[28,29]. Since $N_c$ scales with the nanocrystal volume $V$ (the number of carriers increases as the size increases) and $\alpha$ is determined by the total interband oscillator strength per unit volume $f_{osc}(R)/V$, Eq. (10) can be approximated by

$$\overline{I_{PL}}(\lambda) \cong \int_0^\infty f_{osc}(R)I_{PL}^{(R)}(\lambda)P(R)dR$$
$$= f_{osc}(R)P(R)\frac{1}{\lambda'(R)}\int I_{PL}^{(R)}(\lambda)d\lambda \quad (11)$$

In Eq. (11), the fluorescence lineshape for a fixed radius, $I_{PL}^{(R)}(\lambda)$, relates the distributions $\overline{I_{PL}}(\lambda)$ and $P(R)$ whose abscissas are connected by the relation $\lambda(R) = hc/E_g(R)$, so that $d\lambda = d[hc/E_g(R)] = \lambda'(R)dR$, thus allowing the change in the variable of integration. $h$ is the Planck's constant, $c$ is the speed of light and $E_g(R)$ is the nanocrystal bandgap written explicitly as a function of the radius $R$, for a given set of descriptive parameters, as defined in Eqs. (2) and (9). Considering a normalized spectral lineshape (typically, a Gaussian profile), $\int I_{PL}^{(R)}(\lambda)d\lambda = 1$, the experimentally measured $\overline{I_{PL}}(\lambda)$ can be converted into a size distribution $P(R)$ through the relation:

$$P(R) \cong \frac{1}{f_{osc}(R)}\left[\frac{d\lambda}{dR} \times \overline{I_{PL}}(\lambda)\right]_{\lambda=\frac{hc}{E_g(R)}}$$
$$\cong \frac{1}{V}\left[\frac{d\lambda}{dR} \times \overline{I_{PL}}(\lambda)\right]_{\lambda=\frac{hc}{E_g(R)}} \quad (12)$$

In Eq. (12), the total interband oscillator strength, $f_{osc}(R)$, is obtained by integrating over all the optically allowed exciton states. As discussed in references[12,31], the magnitude of $f_{osc}(R)$ is determined by the total interband matrix element $p_{cv}$ between the valence-band top and the conduction-band bottom, and also by the number of unit cells contained in the nanocrystal. Since $p_{cv}$ is defined in terms of the Bloch wave functions of the bulk material, accounting for semiconductor's composition and crystal lattice, which do not depend on the nanocrystal size, it can be expected that $f_{osc}(R)$ scales linearly with the nanocrystal volume $V$. It is worth pointing out that for small nanocrystals where confinement effects are significant and at relatively low temperatures, the first excited eigenstate is situated at much higher energies than the thermal energy $k_B T$. In this picture, the oscillator strengths of all $(R/a_B)^3$ levels are mainly concentrated on the lowest exciton state[32,33] so that the overall $f_{osc}(R)$ is essentially determined by $f_1(R)$. In such situation, the major contribution to luminescence is from radiative recombination of confined ground-state excitons, the thermal broadening (< 50 meV at room temperature) being negligible in comparison to the observed spectral linewidths[33]. As a consequence, photons emitted at a given energy arise basically from nanocrystals whose lowest excited state corresponds to that energy. Therefore, according to Eq. (12), for a given experimental photoluminescence spectrum, $\overline{I_{PL}}\left(\lambda = \frac{hc}{E_g(R)}\right)/V$ represents approximately the volume fraction of nanocrystals with energy bandgap $E_g(R)$ that is converted into a particle size distribution $P(R)$ through the factor $[d\lambda/dR]_{\lambda=hc/E_g(R)}$.

Alternatively, the size distribution can also be obtained from analysis of the inhomogeneous broadening observed in the optical absorption spectra of semiconductor nanocrystals. Pesika et al.[34,35] estimated $P(R)$ from the local slope of the absorption spectrum $A(\lambda)$ in the vicinity of the onset through the relation:

$$P(R) \cong -\frac{1}{V}\frac{dA}{dR} = -\frac{1}{V}\left[\frac{dA}{d\lambda} \times \frac{d\lambda}{dR}\right]_{\lambda=\frac{hc}{E_g(R)}} \quad (13)$$

Using the proposed bandgap equation [Eq. (2)], the results from Eq. (13) will be compared to those obtained from the photoluminescence based size distribution model [Eq. (12)]. As will be shown in section IV, our analytical expression for the nanocrystal bandgap can be used to improve dramatically the size distribution predictions resulting from Brus model [Eq. (9)], enabling a direct comparison with experimental data.

The here presented theoretical models are suitable for describing systems composed of very small semiconductor nanocrystals belonging to the strong confinement limit. As discussed in section II.A, in this situation the nanocrystal radius is much smaller than the exciton Bohr radius ($R \ll a_B$), which allows one to treat the effective electrostatic interaction between charge carriers as a perturbation of the dominant kinetic energy contribution[21]. The bandgap relation [Eq. (2)] required for the size distribution computation [Eqs. (12) and (13)] was obtained in this specific size range, thus establishing a limit for the applicability of the developed analysis. In particular for cadmium telluride (CdTe), the semiconductor material we are interested in, the exciton Bohr radius is $a_B = 7.5$ nm[36].

## III. EXPERIMENTAL

### A. Preparation of the nanocrystals

The nanocrystals were synthesized using the colloidal chemistry approach in which the particle growth occurs in a solution of chemical reagents containing the metallic cation and the anion sources such as a cadmium salt and a suitable chalcogenide precursor. In this wet chemical preparation, organic stabilizing agents are used in order to inhibit the excessive growth of the evolving particles to a bulk macrocrystalline phase. In the present work, two synthetic routes were adopted. Initially, CdTe nanocrystals were produced following a two-step procedure in accordance with previous references[37,38]. In the first step, $NaBH_4$ (3.56 mmol) and tellurium powder (0.59 mmol) were mixed with 10 mL of deionized water in a 25 mL three-necked flask sealed with rubber plugs. Under intense argon flow, the mixture was stirred gently at room temperature and about 3 hours later, a clear purple solution was observed. The generated NaHTe precursor was then transferred carefully into a closed reaction vessel with 100 mL of degassed water. The inert atmosphere was again necessary to store the fresh NaHTe properly and to avoid oxidation. In the second step, 40 mL of the freshly prepared NaHTe solution were injected, under an intense argon flow and vigorous stirring, in a three-necked flask fitted with rubber septa and containing $CdCl_2$ (1.11 mmol), deionized water (125 mL), and thioglycolic acid (2.88 mmol). The pH value of the cadmium precursor solution was adjusted to 11.1 with 1M NaOH solution before injection of NaHTe. Then, the reaction mixture was heated to 100 $^0$C (reflux temperature) for one hour, and a sample was taken for further characterization and theoretical analysis.

Concerning the characterization procedure, all optical measurements were performed at room temperature. UV-visible spectroscopy was carried out with a Shimadzu UV-Vis-1501 spectrophotometer. Photoluminescence was measured using a modular system consisting of a 378 nm light-emitting diode laser (COHERENT CUBE) as the excitation source and an Ocean Optics USB 4000 spectrometer for collecting the PL emission. Atomic force microscopy (AFM) analysis for determination of particle size distribution was conducted using an NT-MDT-NTEGRA Prima multifunctional scanning probe microscope in a tapping mode. Noncontact "golden" silicon cantilevers (NSG10 series/NT-MDT) with a typical resonance frequency of 240 kHz and a spring constant of 11.8 N/m were used. Once the sample was scanned, the particle height distribution was assessed using SPIP$^{TM}$-Analytical Software for Microscopy[39]. For a nearly spherical shape, which is a reasonable assumption for nanocrystals prepared by the described colloidal chemistry methods, the height measurement corresponds to the size or diameter of the nanocrystal[40]. With respect to sample preparation, a micropipette was used to disperse two droplets ($\approx 10\ \mu L$, each one) of the undiluted nanocrystal solution on a freshly cleaved mica substrate. After 15 minutes, the substrate containing the deposited nanocrystal solution was placed in a Petri dish where a careful immersion in deionized water at room temperature took place for 10 minutes. Then, the water was removed and the Petri dish/sample system was slowly dried in a muffle furnace at 80 °C for about one day. After that, the sample was ready for AFM imaging.

CdTe nanocrystals were also produced following a one-pot approach in accordance with reference[41]. Briefly, 0.43 mmol $CdCl_2.H_2O$ was diluted in 80 mL of ultrapure water in a 100 mL Beaker. L-glutathione (0.52 mmol) was added while stirring, followed by adjusting the pH to 10.0 with a solution of 1.0 mol L$^{-1}$ of NaOH. Next, this solution was added in a 100 mL three-neck flask with a reflux column and a thermocouple coupled with a thermal heater (Cole & Parmer®) in order to control the temperature. Then, 0.04 mmol $Na_2TeO_3$ and 1.0 mmol $NaBH_4$ were added to the solution, followed by reflux at 100°C for one hour. After that, the sample was purified by adding acetone for precipitation of the nanoparticles.

Ultraviolet-visible (UV-vis) spectrum was registered on a diode array UV-2550 Shimadzu spectrometer. Fluorescence spectrum (PL) was obtained at room temperature, using a Shimadzu RF-5301 PC spectrofluorophotometer equipped with a xenon lamp of 150W. Transmission electron microscopy (TEM) was performed on a JEM 2100 FEG-TEM operating at 200 kV (LNNano- Brazilian Nanotechnology National Laboratory). Suspensions of CdTe QDs samples were dispersed in 300-mesh Lacey Formvar with an ultrathin carbon film, which was previously treated by Argon plasma to make it hydrophilic. Several images were registered and the size of the nanoparticles was measured using the ImageJ software.

The characterization procedures were described separately for each sample since the reported syntheses were performed in two different research groups.

## IV. RESULTS AND DISCUSSION

Figure 1 displays typical room temperature absorption and photoluminescence spectra of two colloidal CdTe nanocrystal samples obtained from different synthetic methods as described in section III: a two-step procedure that uses thioglycolic acid as stabilizer agent (TGA-capped CdTe nanocrystals) and a one-pot approach based on L-glutathione (GSH-capped CdTe nanocrystals). In Figs. 1(a) and 1(b), the solid and dashed curves correspond to the measured emission and absorption intensities, respectively, for both CdTe/TGA and CdTe/GSH nanocrystal samples. The fits to experimental data comprise, for each sample, the entire PL band (squares) and also the absorption edge (circles), that is, the region extracted from the absorbance spectrum ($A(\lambda)$) ranging from the onset to the point where $d^2A/d\lambda^2 = 0$. The energy corresponding to the absorption onset can be obtained by plotting the linear function $(Ah\nu)^2 = C(h\nu - E_{onset})$ and finding its intercept ($A$ is the absorbance, $h\nu$ is the photon energy and $C$ is a constant). Appropriate fitting functions were chosen in order to reproduce accurately the available experimental data. The data enclosed in the absorbance edge regions were then fitted to four-parameter lognormal functions. PL experimental points were, in turn, fitted to an exponentially modified Gaussian function (CdTe/TGA sample) and to an asymmetric double sigmoidal function (CdTe/GSH sample). From these fitting functions, the measured photoluminescence and absorption intensities ($\overline{I_{PL}}(\lambda)$ and $A(\lambda)$ in Eqs. (12) and (13), respectively) are written explicitly as functions of $\lambda$. Once the theoretical sizing curve $\lambda = \lambda(R) = \frac{hc}{E_g(R)}$ is determined, the size distribution curves $P(R)$ can be estimated for the analyzed samples. The two different approaches used for the nanocrystal bandgap $E_g(R)$ are represented by Eqs. (2) and (9).

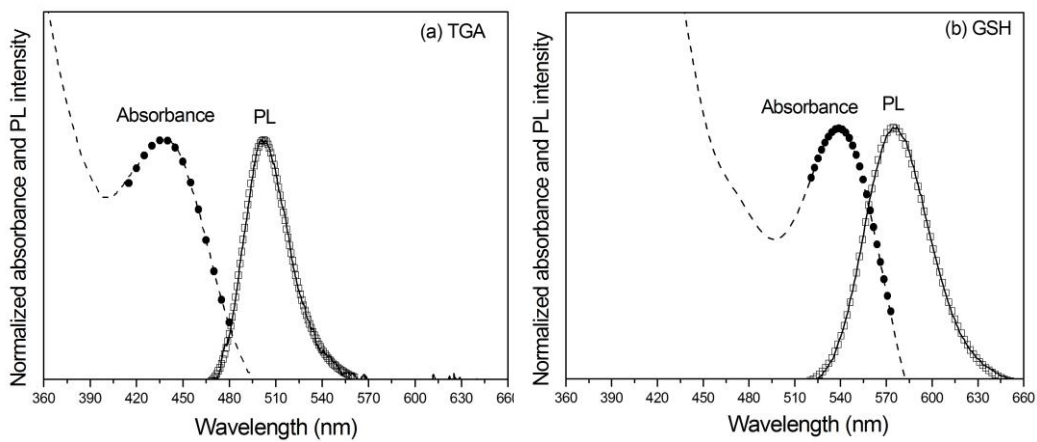

**Figure. 1**. UV-visible absorption and photoluminescence spectra of as-prepared CdTe colloidal nanocrystals capped with (a) thioglycolic acid (TGA) and (b) L-glutathione (GSH). Circles and squares represent fits to absorbance (dashed curves) and PL (solid lines) experimental data, respectively.

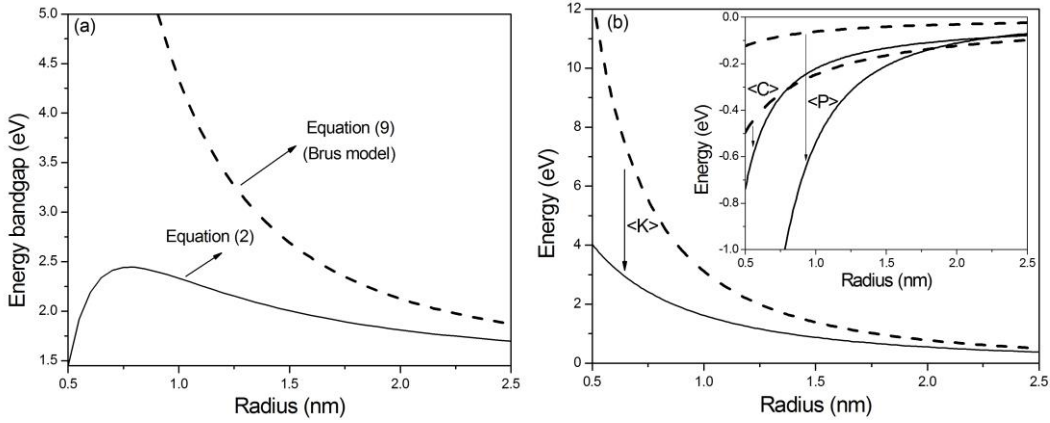

**Figure 2.** (a) Calculated bandgap $E_g(R)$ for a CdTe colloidal nanocrystal in aqueous solution through Eq. (2) (solid line) and from Brus model [Eq. (9), dashed line]. (b) Decomposition of both $E_g(R)$ curves into $\langle K \rangle$, $\langle C \rangle$ and $\langle P \rangle$ contributions (kinetic, coulomb and polarization energies). The arrows connect the quantities calculated in the infinite barrier model [Eq. (9), dashed lines] and in our theoretical approach [Eqs. (2), (6) and (8), solid lines].

In what follows, our theoretical results are presented and compared to the predictions of Brus model [Eq. (9)]. Figure 2(a) shows the size dependent bandgap $E_g(R)$ estimated from Eqs. (2), (6) and (8) for CdTe colloidal nanocrystals in aqueous solution (solid line). The calculations were performed with the parameters: $E_g^{bulk} = 1.475$ eV, $m_e = 0.135 m_0$, $m_h = 1.139 m_0$ ($m_0$ is the free electron mass), and $\varepsilon_s = \varepsilon_{CdTe} = 10.4$. Since the analyzed nanocrystal samples were produced by means of purely aqueous medium routes (section III), the following values of dielectric mismatch and potential barrier height were used: $\varepsilon = \varepsilon_{CdTe}/\varepsilon_{water} = 0.13$ and $V = [E_g^{(water)} - E_g^{(CdTe)}]/2 = 2.7125$ eV; $E_g^{(water)} = 6.9$ eV is the experimental bandgap of liquid water. With the theoretical considerations here proposed, a strong reduction of the nanocrystal bandgap values predicted by Brus model [Eq. (9), dashed curve] is observed in a small size range ($R < 2.5$ nm). Furthermore, as a consequence of the incomplete confinement of the carriers (due to a finite $V$), a clearly noticeable inflexion point in the $E_g(R)$ continuous curve indicates an onset for the vanishing of the bound states in the finite spherically symmetric well (for $R \leq 0.72$ nm, the exciton is no longer confined). In Figure 2(b), the individual contributions of the expectation values of the kinetic energy [$\langle K \rangle$, second and third terms in Eq. (2)], the Coulomb energy [$\langle C \rangle$, Eq. (6)] and the polarization energy [$\langle P \rangle$, Eq. (8)] to the total $E_g(R)$ curve are presented separately (solid lines, our results) and compared to the corresponding predictions of Brus model (dashed lines). The arrows indicate how $\langle K \rangle$, $\langle C \rangle$ and $\langle P \rangle$ change after implementation of the analytical corrections derived in section II. The reduction in the dominant kinetic energy contribution becomes quite large in the strong confinement region, and an attenuated size dependence is verified for $\langle K \rangle$: as the nanocrystal size decreases, $\langle K \rangle$ increases as $R^{-1.4}$ instead of $R^{-2.0}$ (the typical quantum localization term in Eq. (9) scales with the square of the inverse radius). On the other hand, the contribution of $\langle P \rangle$ to $E_g(R)$ is greatly enhanced for small sizes (inset), which is mainly attributed to a significant spreading of the electron and hole probability densities outside the nanocrystal by relaxing the hard-wall boundary condition, as discussed in reference[24]. In fact, while polarization energy is supposed to shift $E_g(R)$ to lower energies as $R^{-1.0}$ (see Eq. (9)), a stronger size dependence was obtained: in our calculations, $\langle P \rangle$ scales with $R^{-2.6}$. Therefore, at small values of $R$, $\langle P \rangle$ becomes much more negative than expected from the infinite barrier model in which polarization effects seems to be almost suppressed. It can also be observed that the magnitude of the Coulomb energy $\langle C \rangle$ in the observed size range is not significantly affected by the existence of a finite confinement potential (inset), which is partly due to the long-range character of the Coulomb interaction. In particular for $R = 0.72$ nm, $\langle K \rangle$ changes from 6.01 eV to 2.55 eV, $\langle C \rangle$ from -0.34 eV to -0.37 eV, $\langle P \rangle$ from -0.09 eV to -1.23 eV, and the calculated bandgap is drastically reduced from 7.14 eV to 2.43 eV.

In Figure 3(a), the particle size distributions (PSDs) obtained from the analysis of both the emission and the absorption spectra for the produced CdTe/TGA nanocrystal sample [Fig. 1(a)] are superimposed on the distribution obtained from the analysis of the displayed AFM image [Fig. 3(b)]. The photoluminescence based PSD (solid line) calculated directly from Eq. (12) and the proposed relation for $E_g(R)$ [Eq. (2)] exhibits a clear asymmetric shape with a most probable radius of 0.82 nm in close agreement with the AFM histogram (white bars with a maximum height centered at 0.81 nm). Such agreement arises from the theoretical considerations that led to a general expression for the nanocrystal bandgap in the form of Eq. (2). Even for nanocrystals embedded in liquid mixtures, the incompleteness of the confinement must be considered as a relevant aspect that affects the different energetic contributions (kinetic, coulomb and polarization energies) to the effective bandgap which, in turn, is greatly reduced in very small nanocrystals. For example, the bandgap corresponding to $R = 0.82$ nm is reduced from 5.8 eV to 2.4 eV when the corrections enclosed in each term of Eq. (2) are implemented. As discussed in section II, in a situation in which the dimensionless confining parameters $v_{i=e,h}$ are considered ideally high, all terms of Brus equation are asymptotically recovered. Consequently, the mechanism of bandgap reduction presented in Fig. 2 is no longer assessed, and the PSD will be dislocated to larger radii. The inset in Fig. 3(a) shows the photoluminescence based PSD [Eq. (12)], using now the Brus approximation to $E_g(R)$ [see Eq. (9)]. The most probable radius is, in fact, strongly overestimated (PSD maximum centered at 1.64 nm). Making use of our bandgap relation again [Eq. (2)], the PSD corresponding to the dashed curve was calculated from Eq. (12), as before, but a different approach was used for the oscillator strength of the lowest exciton state, $f_1(R)$. Since we are dealing with extremely small particles, $R/a_B \cong 0.1$ for the most probable radius, it seems reasonable that $f_1(R)$ can also be represented by an asymptotic limit analogous to that proposed by Kayanuma[31]: for $R/a_B \to 0$, the normalized oscillator strength of the ground state per nanocrystal tends to $f_1^n(R) = \pi|\theta(v_e, v_h)|^2$, where $\theta(v_e, v_h) = \int \phi_{v_e}(x)\phi_{v_h}(x)d^3x$ is the overlap integral calculated from wave function (1) (see section II). For infinite confining potentials $(v_{e,h} \to \infty)$, the classical Kayanuma result for the strong confinement limit is recovered, that is, $f_1^n(R) \to \pi$. As before, the calculated PSD presents an accurate estimate for the most probable radius. Furthermore, the observed asymmetric shape becomes noticeably broader to the right of the maximum in clear agreement with the AFM statistical data. Finally, the PSD obtained from the analysis of the absorbance spectrum in Fig. 1(a) is represented by the grey filled curve. This is the result from the implementation of Pesika model[34,35] [Eq. (13)] combined with our bandgap equation [Eq. (2)]. Although the absorption based PSD furnishes a good estimate for the most probable radius (0.79 nm), the distribution is highly symmetrical and much sharper than those obtained from analysis of both the AFM image and the photoluminescence spectrum. Such discrepancy is inherent to the basic assumption underlying Eq. (13). If the particle size distribution is sufficiently large, then the shape of the absorbance spectrum near the onset is dominated by the particle size distribution. In this situation, the analysis of the absorption edge led to better results for CdSe colloidal nanocrystals produced at prolonged reaction times after a natural broadening of the absorption bands with time[42]. In an opposite situation, our CdTe/TGA nanocrystal sample (corresponding to a short reaction time) exhibits a relatively narrow well-resolved absorption peak, which limits the analysis of size distributions from the absorption spectrum through Eq. (13).

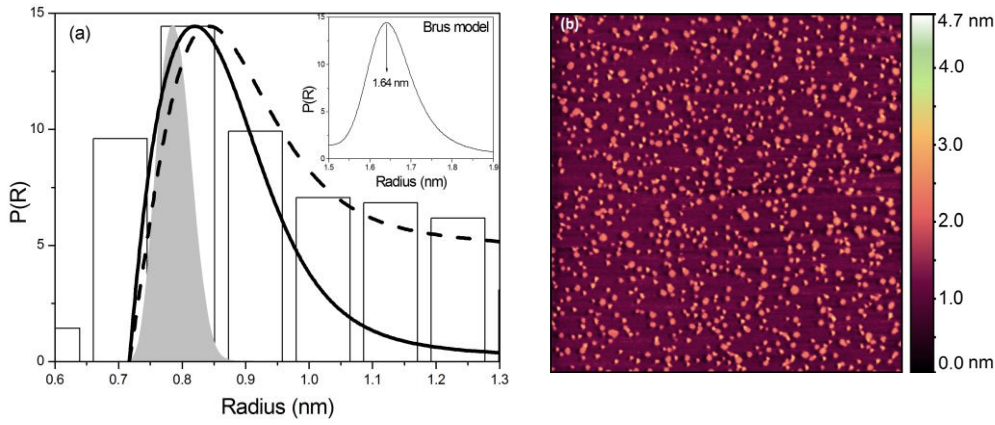

**Figure 3.** (a) Size distributions of CdTe/TGA nanocrystals obtained from: AFM histogram (white bars); absorption edge with $E_g(R)$ given by Eq. (2) (grey filled curve); PL spectrum with $E_g(R)$ given by Eq. (2) (solid line); PL spectrum with $E_g(R)$ given by Eq. (2) and making use of an asymptotic oscillator strength term (dashed curve); PL spectrum with Brus approximation to $E_g(R)$ [Eq. (9), inset]. (b) AFM image of CdTe/TGA nanocrystal sample (height distribution).

In Figure 4(a), similar analyses were performed for the produced CdTe/GSH nanocrystal sample using the emission and the absorption spectra displayed in Fig. 1(b). The PL based PSD (solid line) calculated from Eqs. (2) and (12) exhibits an asymmetric shape in close agreement with the size distribution histogram (white bars) obtained from the corresponding TEM image [Fig. 4(b)] except for the region in the vicinity of $R = 1.0$ nm where experimental results are noticeably underestimated. This may be the result of the difficulty in obtaining precise measurements of smaller particles from TEM images[42]. Since we are dealing with a system characterized by a considerable size dispersion (27%), this may also indicate that a post preparative procedure such as the size selective precipitation technique[26,27] should be used conveniently to produce new samples with narrower size distributions before the PSD computation. In the infinite potential limit $v_i \to \infty$, the calculated distribution centered at 1.25 nm (solid line) shifts to 1.98 nm (inset, Brus model) far from the TEM statistical data, as a consequence of the use of the asymptotic formula (9) for the nanocrystal bandgap. The grey filled curve represents an absorbance based PSD with an approximately symmetric shape computed from Eq. (13) and our expression for $E_g(R)$ [Eq. (2)]. This distribution (centered at 1.21 nm) is much sharper than those obtained from analysis of photoluminescence spectrum and TEM data for the same reasons discussed previously. Since the size distribution of the CdTe/GSH sample is dislocated substantially to the right of the distribution of the CdTe/TGA sample, the previously investigated asymptotic limit for the oscillator strength turned out to be unsuitable to describe the larger particles in the GSH-capped nanocrystal sample. In fact, the CdTe/GSH sample exhibits broader absorption and emission bands situated at much longer wavelengths than the CdTe/TGA sample (Fig. 1).

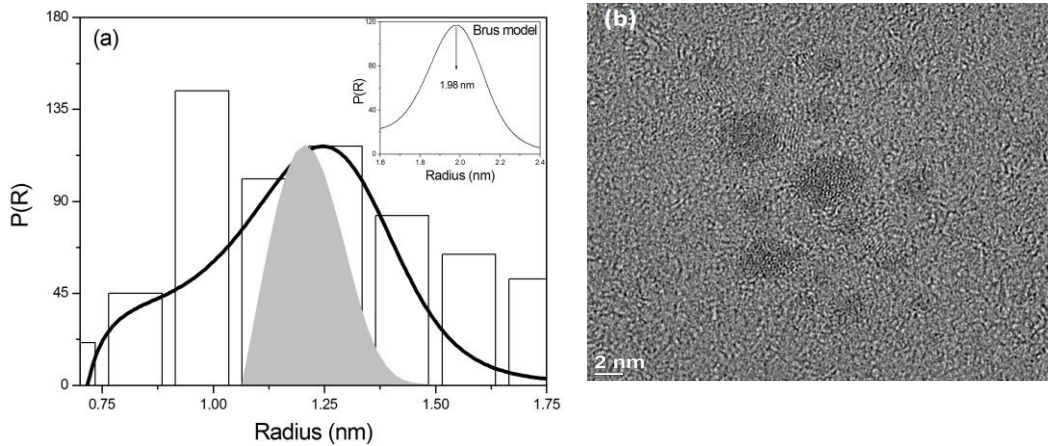

**Figure 4.** (a) Size distributions of CdTe/GSH nanocrystals obtained from: TEM histogram (white bars); absorption edge with $E_g(R)$ given by Eq. (2) (grey filled curve); PL spectrum with $E_g(R)$ given by Eq. (2) (solid line); PL spectrum with Brus approximation to $E_g(R)$ [Eq. (9), inset]. (b) TEM image of CdTe/GSH nanocrystal sample.

The influence of different effective mass values on PSD computation was analyzed in terms of the anisotropy effect in zinc-blende (bulk) semiconductor materials. Such effect is more pronounced for the heavy-hole band which has a strongly directional-dependent effective mass, with a larger mass along the [111] direction than along the [110] and [100] directions[43]. Using a theoretical methodology developed for the first author of the present paper[44] and here applied to the binary semiconductor CdTe, the electron and the heavy-hole effective masses along these three directions were determined from *ab initio* total energy calculations based on the density functional theory[45]. Then, mean effective masses were obtained by averaging over the directions. In Figs. 2, 3 and 4, all calculations were performed with the [111] effective masses that led to the most accurate descriptions of the measured distributions. It is worth pointing out that, in the observed size range of the analyzed samples, corresponding to the strong confinement regime, the agreement between theoretical predictions and experimental data was little affected when the mean effective masses were used in the calculations. Indeed, the PSDs estimated from these two sets of parameters ([111] effective masses and mean effective masses) are quite similar and exhibit very close values for the most probable sizes. However, when the effective masses along the [110] and [100] directions were used, the most probable sizes increased significantly with respect to the values corresponding to the first two sets of parameters, compromising the comparison with experimental data. These considerations are equally valid for several zinc-blende binary semiconductor materials (CdS, CdSe, ZnS, ZnTe, ZnTe, and others) and must be taken into account in order to determine the size distribution of ensembles of nanocrystals properly. For the sake of completeness, the calculated electron and heavy-hole effective masses are listed here:

$m_e^{[100]} = 0.131 m_0$, $m_{hh}^{[100]} = 0.506 m_0$,
$m_e^{[110]} = 0.133 m_0$, $m_{hh}^{[110]} = 0.520 m_0$,
$m_e^{[111]} = 0.135 m_0$, $m_{hh}^{[111]} = 1.139 m_0$,
$m_e^{Mean} = 0.133 m_0$, $m_{hh}^{Mean} = 0.825 m_0$.

In order to summarize the main ideas proposed in this paper, a schematic diagram showing our general approach to size distribution determination is presented in Figure 5 (steps 1 to 7). For a particular system of semiconductor nanocrystals embedded in a specific medium, a set of descriptive parameters is initially defined (1): bandgap values of the bulk semiconductor material ($E_g^{bulk}$) and of the surrounding medium ($E_g^{medium}$), dielectric mismatch ($\varepsilon$), effective masses of the confined charge carriers ($m_e, m_h$), and barrier height ($V$). These initial parameters are used to calculate the nanocrystal bandgap (2) which, in turn, allows one to convert PL [$\overline{I_{PL}}(\lambda)$] and aborbance [$A(\lambda)$] data into size distribution curves (3). $\overline{I_{PL}}(\lambda)$ and $A(\lambda)$ are obtained from optical measurements previously performed on suspensions of as-prepared colloidal nanocrystals (4,5). Subsequent AFM/TEM characterization (6) yields the particle size distribution histogram, enabling a direct comparison with theoretical predictions (7).

## V. CONCLUSIONS

In the present work, we have calculated the size-dependent bandgap of colloidal semiconductor nanocrystals from an extensive revision of the main theoretical contributions to the understanding of this well-known quantum confinement effect.

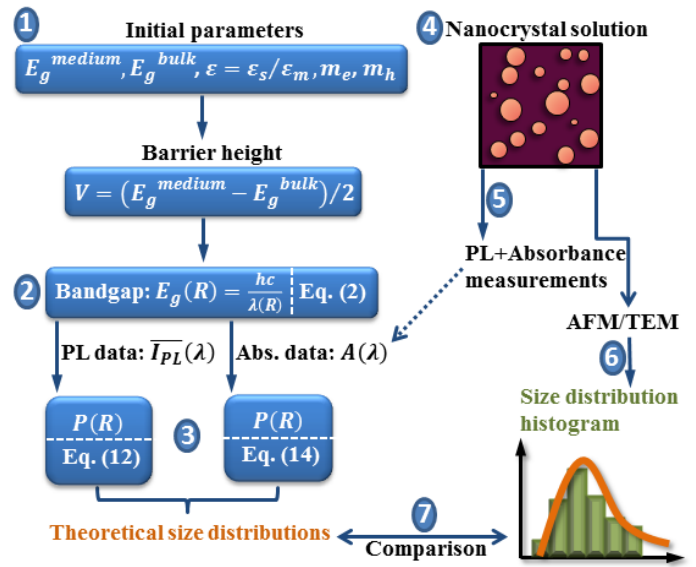

**Figure 5.** Schematic diagram showing in a few steps (1 to 7) the method employed to determine particle size distribution through spectroscopic data.

By considering the exact wave function for the charge carriers confined in a finite spherical potential well, the relevance of the incompleteness of the confinement can be quantified. Once finite confining potentials are considered, the expectation values of the kinetic energy, the electron-hole Coulomb interaction and the polarization energy are calculated properly leading to a dramatic reduction of the nanocrystal bandgap. Consequently, the so-called inadequacy of the effective mass approximation for small nanocrystal sizes is overcome. In fact, the size distributions obtained from the analysis of the photoluminescence spectrum together with the proposed bandgap equation are directly comparable to the presented AFM and TEM data. Precise estimates for the most probable radius were provided as well as relatively broad and asymmetric shapes in close resemblance to the measured distributions. On the other hand, the particle size distributions obtained from the most common analysis of the absorbance edge turned out to be almost symmetrical and much narrower than the measured distributions as already discussed in other publications. The methodology presented is this paper for bandgap calculation and particle size determination can be easily implemented and extended to other systems of semiconductor nanocrystals. It can be used as a complementary tool for the characterization of ensembles of nanocrystals produced from different synthetic approaches. Finally, the possibility of recovering the size distribution from spectroscopic experiments can be used to clarify the growth kinetics of colloidal nanocrystals since the temporal evolution of optical spectra is easily monitored during a typical growth experiment. The growth kinetics of TGA-capped CdTe nanocrystals was completely described by the present authors in the sense of the classical crystallization theories by employing this methodology. These new results will be shown in a forthcoming publication.

## ACKNOWLEDGMENTS

This work was supported by the Brazilian agencies CAPES, CNPq, and FAPEMIG.